\documentclass[doublecol]{epl2}

\title{Comment on "On quantum plasma: A plea for a common sense [Europhys. Lett., 99 25001 (2012)]"}
\shorttitle{Title} 
\author{P. K. Shukla$^{1,2}$ and M. Akbari-Moghanjoughi$^3$}
\shortauthor{P. K. Shukla and M. Akbari-Moghanjoughi}
\institute{
  \inst{1} International Centre for Advanced Studies in Physical Sciences and Institute for Theoretical Physics,
Faculty of Physics \& Astronomy, Ruhr-University Bochum, D-44780 Bochum, Germany\\
\inst{2} Department of Mechanical \& Aerospace Engineering, University of California San Diego, La Jolla, CA 92093\\
\inst{3} Azarbaijan University of Shahid Madani, Faculty of Science, Department of Physics, 51745-406, Tabriz, Iran
}
\begin{abstract}
{We point out flaws in the work of Vranjes {\it et al.} (EPL {\bf 99} 25001, 20012), and present correct
criteria for quantum plasmas.}
\end{abstract}


\begin{document}

\maketitle
In a recent letter, Vranjes {\it et al.} \cite{vranjes} are giving the impression that numerous theories for quantum plasmas are new,
which have been flourishing only during the past few years. This is misnomer, since the importance of the degeneracy of electrons
was recognized through the pioneering works of Fowler and Chandrasekhar \cite{chandra0}. During the early phase of quantum
mechanics, Madelung has developed the quantum fluid theory for electrons by using the Schr\"odinger equation. About sixty years
ago, many distinguished physicists \cite{silin,bohm} laid down the foundation to collective interactions in dense quantum plasmas.
In fact, an early experiment \cite{watanabe} in 1956 had already provided an experimental evidence for the quantum feature of the electron
plasma waves in solids, where the electrons are in a degenerate state, and thereby supporting the linear theory of Klimontovich
and Silin \cite{silin} and  Bohm and Pines \cite{bohm}. Recently, there has been a surge in investigating numerous
nonlinear processes \cite{shukla10, brodin, haas, vlad} in quantum plasmas, where the electron degeneracy
and quantum effects are shown to play a crucial role. Furthermore, applications of collective interactions in quantum plasmas
rest on exploring new physics of stimulated scattering of intense laser beams off quantum electron plasma oscillations \cite{glen}
and quantum free-electron lasers in the x-ray regime, and the discovery of novel attractive force \cite{shukla12} that can
bring ions closer at atomic dimensions in order for the magneto-inertial confinement fusion to work in a highly compressed
high-energy solid density plasma.

Our objective here is to demonstrate that Vranjes {\it et al}'s criteria for the applicability of quantum plasmas are totally
misleading and invalid/erroneous. First of all, we recall that the plasma state occurs in the form of an ionized gas through
which electricity can flow. The Saha's ionization criterion dictates that in order for the plasma to form one must have sufficiently
high degree of ionization. Thus, ordinary air at room temperature is not in the plasma state, since the fractional ionization
$n_i/n_n$ is ridiculously low $\sim 10^{-122}$. The fractional ionization, however, increases with the increase of the gas
temperature by heating, so that the electrons are liberated from neutral atoms and  they are a component of the electron-ion
plasma which can be fully or partially ionized. In a fully ionized classical electron-ion plasma, the thermal de Broglie wavelength
$\lambda_B = \hbar/\sqrt{2 \pi m_e k_B T_p}$, which characterizes the spatial extension of the probability density
of the electrons, is much less than the average inter-electron distance $d \sim (3/4\pi n_0)^{1/3}$,
where $\hbar$ is Planck's constant divided by $2\pi$, $m_e$ the electron mass, $k_B$ the Boltzmann constant, $T_p$ the plasma
temperature, and $n_0$ the average electron number density. In low-temperature laboratory plasmas with  $T_p = 300$ degrees Kelvin,
we then have $n_0 \ll 10^{18} cm^{-3}$, and there is no need to consider the electron degeneracy effect, and one uses
the classical theory for ideal plasmas with the Maxwell-Boltzmann distribution for non-degenerate electrons and ions. However, the
situation is different for metallic conduction electrons where (due to their high mobility) they are degenerate with moderate
density of $3 \times 10^{22} cm^{-3}$ at room temperature. Thus, in order for the quantum effect to become important, the thermal de Broglie
wavelength of quantum particles must be $\leq d$, and the Landau length $\lambda_l= e^2/k_B T_p \geq d $, with $e$ being the
magnitude of the electron charge. It turns out that, $d \geq \hbar/\sqrt{2\pi m_e k_B T_P}$ and $k_B T_p \geq  e^2/d$ are the
appropriate criteria to define the range of a quantum plasma with degenerate electrons.

It must be stressed that in a quantum plasma, degenerate electrons follow the Fermi-Dirac distribution function, which gives
an expression that relates the Fermi electron energy $E_F =k_B T_F$ and $n_0$. One has \cite{shukla10}
$T_F =(\hbar^2/2 k_B m_e)(3 \pi^2 n_0)^{2/3}$, which reflects the quantum nature of the Fermi electron temperature $T_F$ through
$\hbar$. Vranjes {\it et al.} \cite{vranjes} have confused the Fermi-temperature, $T_F$, with the plasma temperature $T_p$. About five years
ago, Glenzer {\it et al.} \cite{glen} reported observations of the electron plasma oscillations in a solid density plasma
(with the peak electron number density $\sim 3 \times 10^{23} cm^{-3}$ and the equilibrium electron and ion temperatures
of 12 eV ($T_p \sim 1. 4 \times 10^{5}$ degrees Kelvin), which is different from the Fermi electron temperature for metals (of order $10^4-10^5K$), by using collective x-ray scattering techniques. Thus, the experiments of Glenzer {\it et al.} \cite{glen} have
unambiguously demonstrated the quantum dispersive effects associated with quantum statistical pressure and the electron
recoil effects in solid density laboratory plasmas.

On the other hand, in his Nobel Prize winning paper, Chandrasekhar \cite{chandra0} presented the famous pressure law for degenerate electrons, which reads
$P_c=(m_e^4c^5/24\pi^2\hbar^3)[R(2R^2-3)\sqrt{1+R^2} + 3 {\rm sinh}^{-1}R]$, with $R=(P_{Fe}/m_e c)=(n_0/n_c)^{1/3}$ where $P_{Fe}$
is the electron relativistic Fermi-momentum and $n_c =m_e^3c^/3\pi^2\hbar^3 \simeq5.9\times 10^{29} cm^{-3}$. The generalized Chandrasekhar degeneracy pressure $P_c$ reduces to ${P_n} \approx ({\pi ^2}/5){(3/\pi )^{2/3}}({\hbar ^2}/{m_e})n_0^{5/3}$ and $P_u \approx (\pi/4) (3/\pi)^{1/3} \hbar c n_0^{4/3}$ in non-relativistic and ultra-relativistic cases obtained in $R \ll 1$ and $R \gg 1$ limits, respectively. The pressure for degenerate electrons and the plasma temperature are related through $n_0 k_B T_p =P_c$.

\begin{figure}
\onefigure{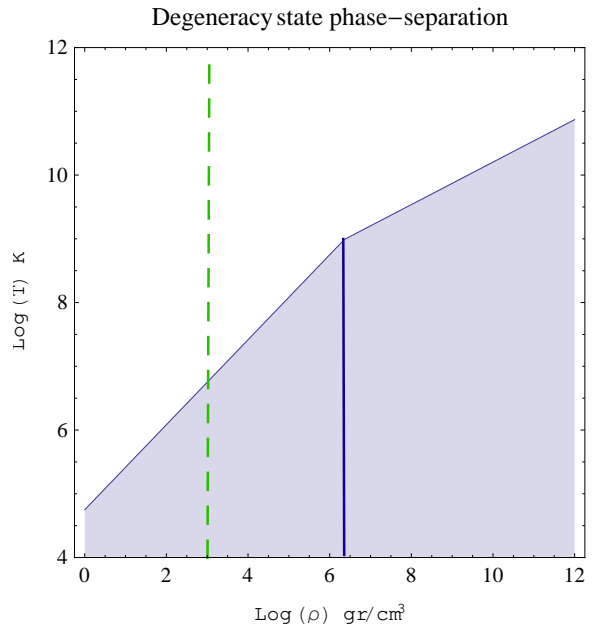}
\caption{Degeneracy phase-separation diagram with the shaded area corresponding to the electron degeneracy.
The vertical dashed line indicates the density above which the plasma is completely pressure ionized.}
\label{fig.1}
\end{figure}

It seems that Vranjes {\it et al.}\cite{vranjes} deduced the electron degeneracy criterion for non-relativistic quantum plasmas from $ n_0 k_B T_F = P_{n}$, which simply reflects the relationship between $T_F$ and $n_0$ and cannot be regarded as correct criterion to define quantum plasmas. The correct criteria for the
electron  degeneracy effect in quantum plasmas come from $d \leq \lambda_B, \lambda_l$, which yield
$n_0^{1/3} \leq (3/4\pi)^{1/3}\sqrt{2\pi m_e k_B T_p}/\hbar, \ (3/4\pi)^{1/3} e^2/k_B T_p$. The latter hold
for quantum plasmas in solids \cite{watanabe}, in high-energy density compressed plasmas \cite{glen}, giant planetry
systems \cite{fortov}, and in  compact astrophysical objects (e.g. white dwarf stars \cite{chandra0}). Finally, we note
that Vranjes {\it et al.}'s \cite{vranjes} have presented an incorrect expression for $T_{deg}$ that is proportional to $n_0^{3/2}$,
and declare it a meanigful criterion for the electron degeneracy in quantum plasmas. To conclude, we can say that
the figure and  the table displayed in Ref. \cite{vranjes} are fallacious, and are of no use for defining the regimes of
quantum plasmas.

Figure 1 exhibits the regions where quantum plasmas are in nonrelativistic and relativistic electron degeneracy states
(the shaded area), which are separated into non-relativistic and relativistic degeneracy regimes at a  critical mass-density
$\log(\rho_{cr})\simeq 6.34$ defined by $P_{n}=P_{u}$. The non-shaded area corresponds to the classical ideal gas phase.

\end{document}